\documentclass[twocolumn,showpacs,amsmath,amssymb,longbibliography,superscriptaddress]{revtex4-2}
\usepackage{color,longtable}
\usepackage[dvips]{graphicx}
\usepackage{amssymb,amsfonts,amsmath}
\usepackage{comment}

\begin{document}%
\title{Fluid--structure coupling governs a dynamic transition in foam scraping}
\author{Rei Kurita}
\author{Masaya Endo}

\affiliation{%
 Department of Physics, Tokyo Metropolitan University, 1-1 Minami-osawa, Hachiouji-shi, Tokyo 192-0397, Japan
}%
\date{\today}

\begin{abstract}
Foam scraping exhibits a dynamic transition between a slip state, in which the foam moves beneath a plate, and a scraping state, in which the foam is expelled from the confined region. 
Although this transition has been associated with the propagation of local T1 rearrangements, the physical parameter controlling their propagation remains unclear. 
Here, we investigate the dependence of the critical scraping velocity on the liquid fraction, bulk viscosity, and surfactant system. 
For all examined conditions, the critical capillary number follows $\mathrm{Ca}_c\propto\phi^{-1}$, apart from a solution-dependent prefactor. 
We propose a local fluid--structure-coupling model in which viscous work transmitted through the Plateau-border network competes with the effective energetic cost required for one T1 event to trigger the next. 
The observed scaling implies an effective energetic cost governed by the Laplace pressure and is consistent with a small local compressive component of the bubble deformation. 
These results identify local coupling between interstitial flow and bubble deformation as a mechanism controlling the macroscopic slip--scraping transition.
\end{abstract}

\keywords{Foam; Scraping; Rearrangement}
\maketitle

\section{Introduction}
Foams are soft jammed materials in which gas bubbles are densely packed within a small amount of liquid. Because the dispersed elements are deformable,
foams belong to a broader class of soft jammed systems that also includes emulsions and biological tissues. Their combination of structural rigidity
and local deformability gives rise to a characteristic set of properties, including elasticity, liquid uptake, thermal insulation, and suppression of mass transport~\cite{Weaire2001,cantat2013}. These properties underlie the widespread use of foams in foods, fire-extinguishing agents, cosmetics, detergents, and insulating
materials. 

In many practical uses of foams, external forces are applied to induce large deformation~\cite{marchand2020, Deblais2015}. 
Examples include controlling food texture, ejecting fire-extinguishing foams, and spreading cosmetic or detergent foams over a surface. 
For such two-component systems, pioneering studies by Tanaka and co-workers have demonstrated that the rheological response can be governed by the dynamical coupling between the dispersed structure and the surrounding fluid~\cite{Araki2008, Furukawa2006, Yanagishima2021, Tateno2025}. 
The mechanical properties of foams, including their yield stress, elasticity, and viscoelastic response, have been studied extensively~\cite{Merrer2012, Furuta2016, Katgert2013, siemens2010, cohen2013}.  
Nevertheless, much less is understood about how the dynamical coupling between the bubble structure and the interstitial liquid selects a macroscopic deformation mode under large-scale forcing.

Recent experiments on foam scraping have revealed non-monotonic deformation patterns that cannot be explained solely by the mean shear rate or the macroscopic yield stress~\cite{Endo2023, Endo2025}. On a hydrophilic substrate, the foam slips beneath the plate at low scraping velocities, whereas at higher velocities it is expelled from beneath the plate, resulting in a scraping state. The critical velocity $U_c$ decreases as the gap distance increases, showing that the transition does not occur at a constant shear rate. 
Furthermore, $U_c$ also depends on the lateral width of the foam, providing further evidence that the transition cannot be described by a simple local shear-rate criterion~\cite{Endo2025}.

The boundary between the slip and scraping states was suggested to exhibit a dynamic critical phenomenon belonging to the directed-percolation universality class~\cite{Sano2016, Haye2000, Takeuchi2009}. In the slip state, a local rearrangement (T1 event) can trigger subsequent rearrangements in the surrounding foam, leading to spatial propagation of T1 activity. However, the physical parameter that determines whether a T1 event propagates, and therefore controls the transition from slip to scraping, remains unknown.

In this study, we show that the critical capillary number for the
slip--scraping transition follows
\begin{equation}
    \mathrm{Ca}_c \sim \phi^{-1},
\end{equation}
where $\phi$ is the liquid fraction. To explain this scaling, we propose a local fluid--structure-coupling model in which the propagation of a T1 event is determined by the balance between viscous work in the Plateau-border network and the energy required for a small local compression of the bubbles near the rearranging region. 
This result identifies the interstitial liquid as an active component that selects the macroscopic deformation mode of the foam. 
The present framework may also provide insight into how local fluid--structure coupling controls macroscopic deformation in other soft jammed systems, including emulsions and biological tissues.

\section{Materials and Methods}
Foams were prepared using aqueous solutions of three surfactant systems:
5.0 wt\% tetradecyltrimethylammonium bromide (TTAB), sodium dodecyl sulfate (SDS), both purchased from FUJIFILM Wako Pure Chemical Corporation, Japan, and a commercial household detergent (Charmy, Lion Corp., Japan).
The TTAB concentration was well above its critical micelle concentration ($\sim0.1$ wt\%)~\cite{Danov2014}. The surface tensions of the TTAB, SDS, and Charmy solutions were measured using a surface tensiometer (DY300, Kyowa Interface Science Co., Ltd., Japan) and were 37, 34, and 25 mN/m, respectively. 
Previous measurements showed that the drainage times of foams prepared with these three surfactant systems were comparable and that their Boussinesq numbers were approximately $\mathrm{Bo} = \eta_s/\eta r \simeq 2$, where $\eta_s$ is an interfacial viscosity, $r$ is the Plateau-border radius, and $\eta$ is a bulk viscosity~\cite{Kaneda2025}. 
Thus, their interfaces have comparable mechanical properties.
The bulk viscosity was varied by adding glycerol to the surfactant solution. 
Within the experimental uncertainty of approximately 0.1 mN/m, the surface tension was independent of the glycerol concentration. 
The equilibrium contact angle on an acrylic substrate was $\theta_E = 23\pm2^\circ$ and was also independent of the glycerol concentration.

Foams were generated using a foam dispenser manufactured by Awahour (Torigoe Plastic Industry Co., Ltd., Japan). 
For each experiment, a portion of the prepared foam was sampled, and its liquid fraction $\phi$ was determined from $\phi = m/\rho V_{\mathrm{sample}}$, where $m$ is the mass of the foam sample, $\rho$ is the density of the liquid phase, and $V_{\mathrm{sample}}$ is the total volume of the foam sample. 
The uncertainty in the measured liquid fraction was approximately $\pm1\%$.
The mean bubble radius was $R = 0.22$ mm, with a standard deviation of 0.082 mm. 
No appreciable drainage was observed during the experimental period, which was at most 20 s.

\begin{figure}[htbp]
\begin{center}
\includegraphics[width=8.5cm]{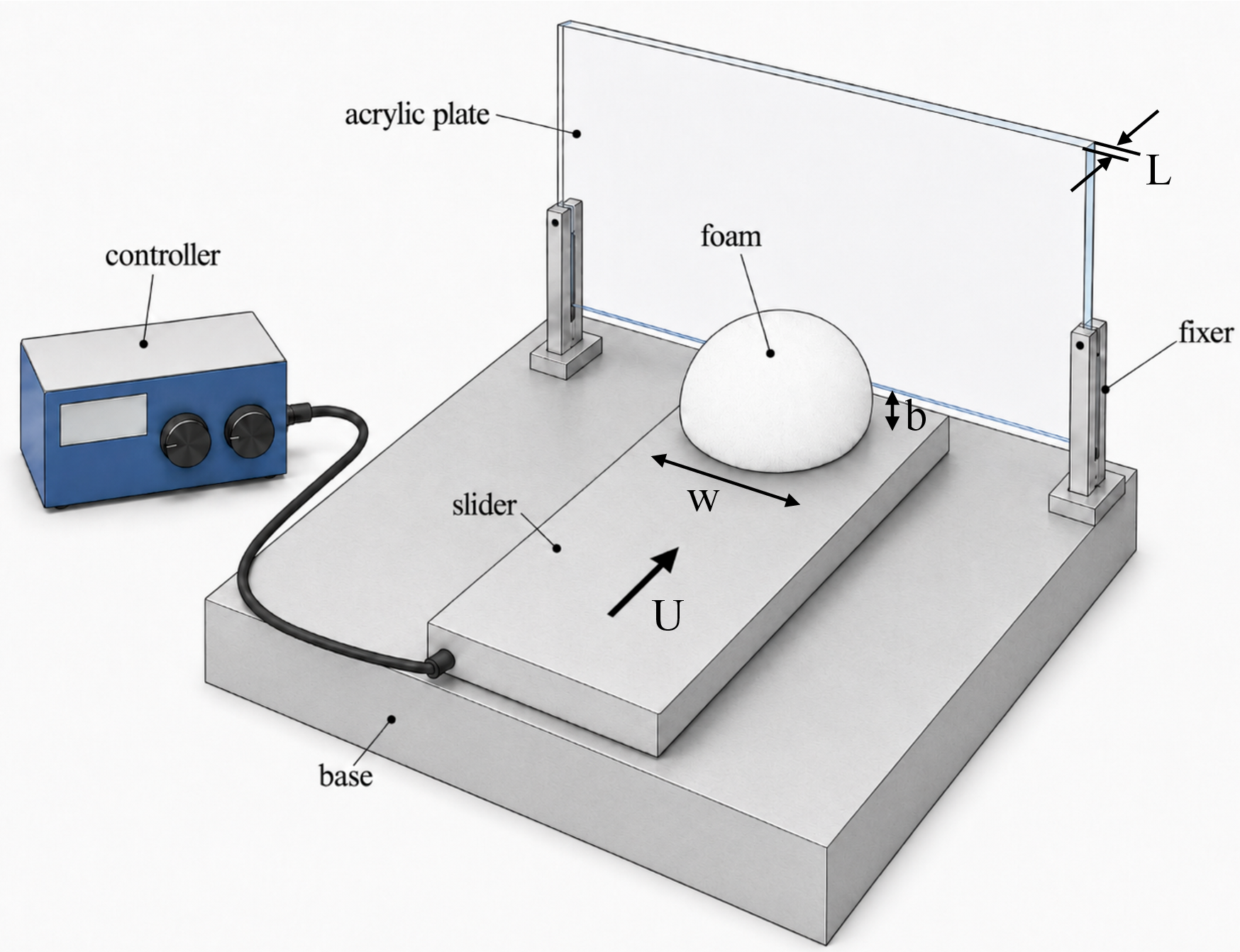}
\caption{
Schematic of the experimental setup for foam scraping.
An acrylic plate was positioned perpendicular to the substrate, forming a gap of width $b = 1.5$~mm and a confinement length $L = 5.0$~mm.
A hemispherical foam sample with a diameter $W = 20$ mm was placed on the substrate, which was translated horizontally at a constant velocity $U$.
}
\label{setup}
\end{center}
\end{figure}

The experimental setup is shown in Fig.~\ref{setup}. 
An acrylic plate was positioned perpendicular to the substrate, forming a gap of width $b = 1.5$ mm. 
The confinement length was set to $L = 5.0$ mm. To construct the foam-scraping state diagram, a hemispherical foam sample with a diameter $W = 20$ mm was placed on the substrate. The substrate was translated horizontally at a constant velocity in the range $U = 1.0$--$35.0$ mm/s using an electric slider (LTS300/M, Thorlabs, Inc., USA). The scraping dynamics were recorded from above using a video camera (EOS R, Canon Inc., Japan).

\section{Results and Discussion}
\subsection{Liquid-fraction dependence of the transition}
The transition from slip to scraping can be described in terms of a directed percolation process, in which an isolated T1 event propagates to neighboring regions and develops into a continuous sequence of rearrangements through the foam~\cite{Endo2023,Endo2025}. 
Within this framework, the critical point is determined by the probability that one T1 event triggers the next.

\begin{figure}[htbp]
\begin{center}
\includegraphics[width=8.5cm]{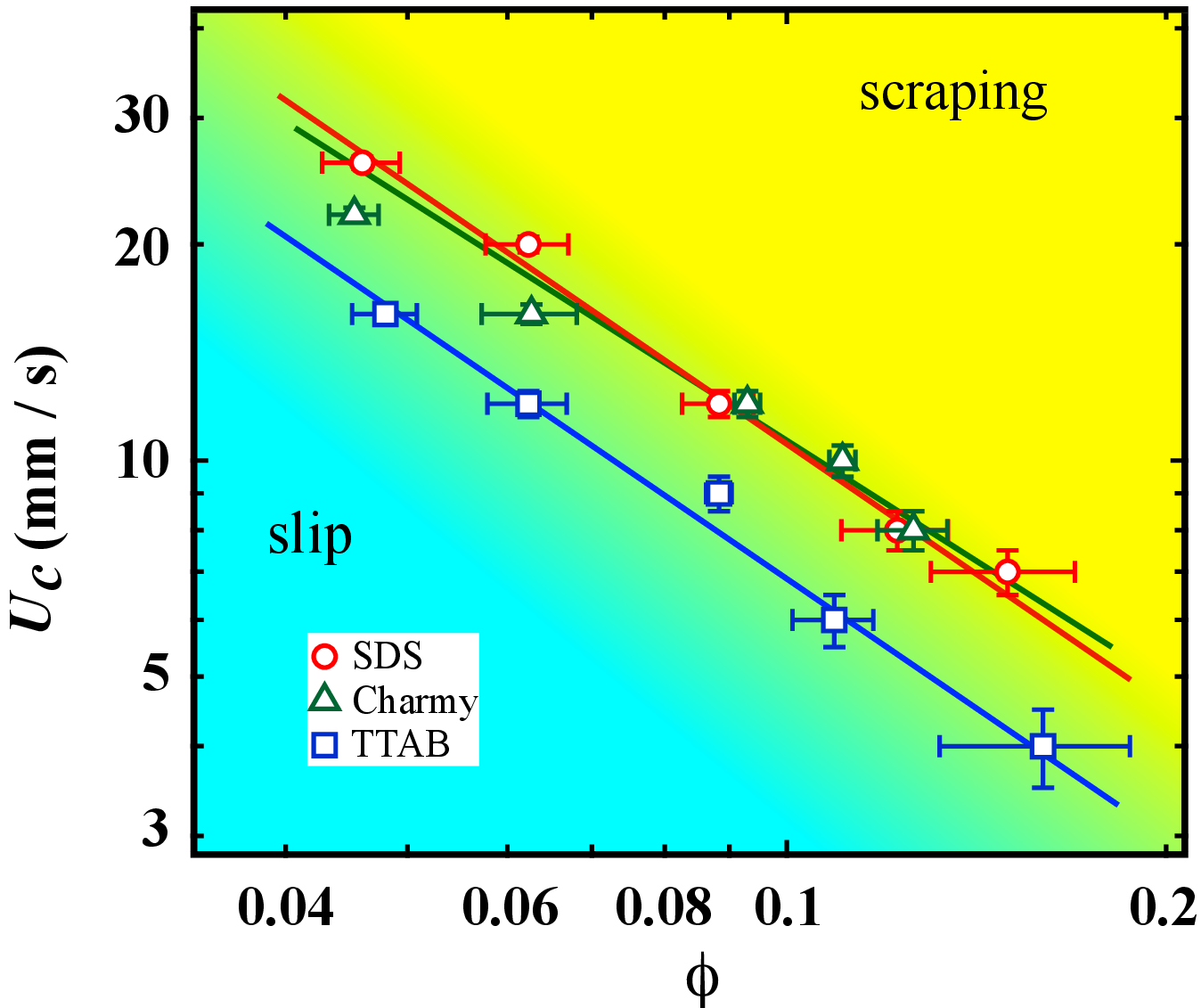}
\caption{
Liquid-fraction dependence of the critical scraping velocity $U_c$ for foams prepared with SDS, Charmy, and TTAB.
The solid lines represent fits to $U_c\propto\phi^{-\alpha}$.
The fitted exponents are
$\alpha = 1.17\pm0.22$ for SDS,
$\alpha = 0.93\pm0.11$ for Charmy, and
$\alpha = 1.16\pm0.18$ for TTAB.
The background colors indicate the slip and scraping regimes.
Horizontal error bars represent the uncertainty in the measured liquid fraction, 
whereas vertical error bars indicate the velocity increment of $0.5$~mm/s used to determine the transition.
}
\label{Uc}
\end{center}
\end{figure}

In foams, the liquid fraction is known to be a key parameter controlling the occurrence of T1 events~\cite{Furuta2016,Kurita2017}. 
We therefore examined the liquid-fraction dependence of the critical scraping velocity $U_c$. 
The results for foams prepared with SDS, Charmy, and TTAB are shown in Fig.~\ref{Uc}. 
For all three surfactant systems, the transition occurred at velocities of the same order of magnitude, and the data were well described by $U_c \propto \phi^{-\alpha}$.
The fitted exponents were $\alpha = 1.17 \pm 0.22$ for SDS, $\alpha = 0.93 \pm 0.11$ for Charmy, and $\alpha = 1.16 \pm 0.18$ for TTAB.
The uncertainties mainly originate from variations in the liquid fraction among independently prepared foam samples. 
In all three systems, the fitted exponent is consistent with $\alpha=1$, indicating
\begin{equation}
    U_c \propto \phi^{-1}.
\end{equation}

\subsection{Capillary-number scaling}
The capillary number, $\mathrm{Ca} = \eta U/\gamma$ is a fundamental dimensionless parameter for describing foam spreading.
To examine the viscosity dependence of the slip--scraping transition, we used the three surfactant systems and additionally prepared TTAB solutions containing
glycerol.

\begin{figure}[htbp]
\begin{center}
\includegraphics[width=8.5cm]{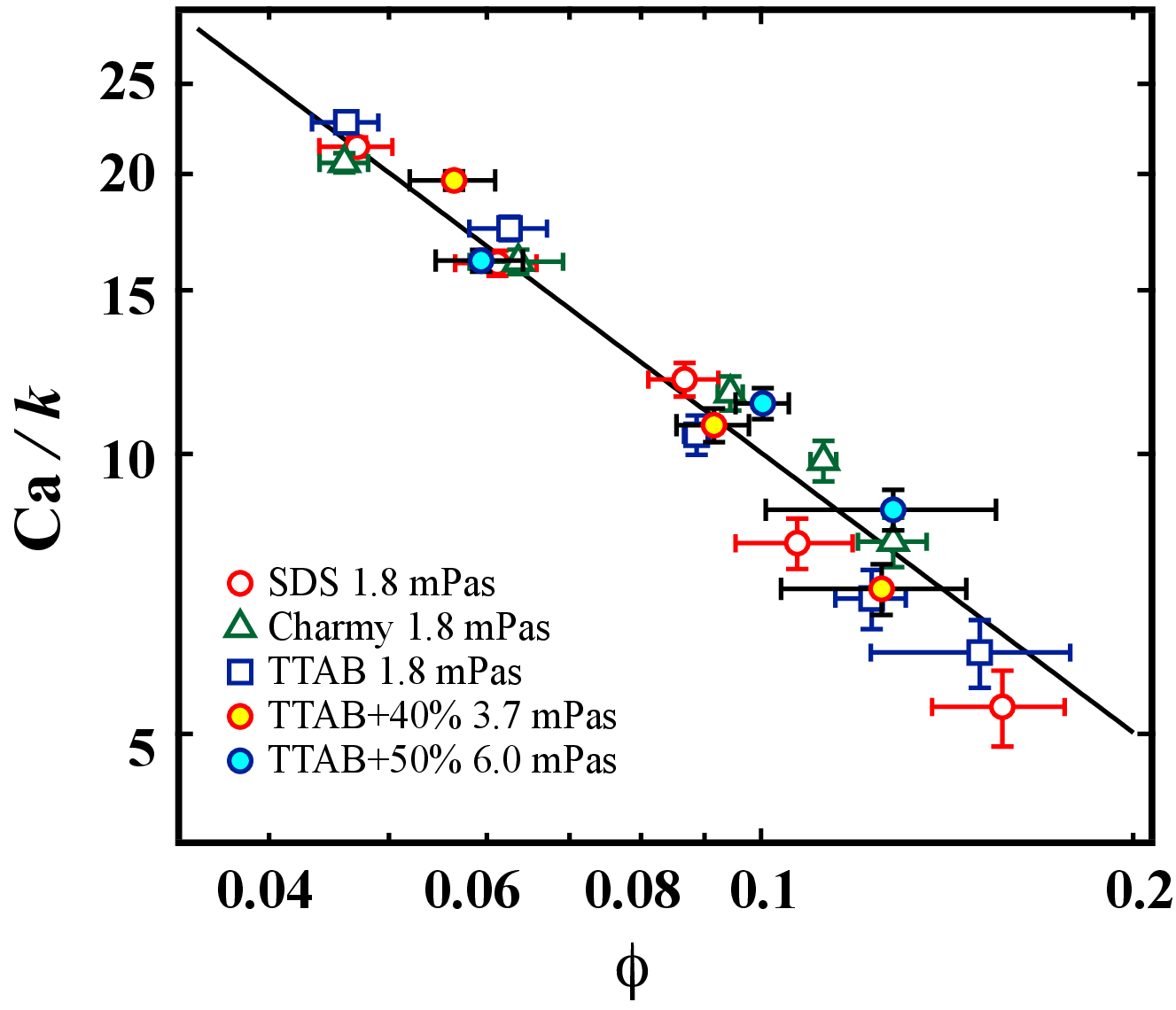}
\caption{
Liquid-fraction dependence of the critical capillary number normalized by the solution-dependent prefactor $k$.
Results are shown for SDS, Charmy, TTAB, and TTAB solutions containing 40 and 50 wt\% glycerol.
The solid line represents $\mathrm{Ca}_c/k=\phi^{-1}$.
Horizontal error bars represent the uncertainty in the measured liquid fraction, whereas vertical error bars are calculated from the uncertainty in the critical velocity.
}
\label{Ca}
\end{center}
\end{figure}

Figure~\ref{Ca} shows the liquid-fraction dependence of the critical capillary number normalized by the solution-dependent prefactor $k$.
After normalization by a solution-dependent prefactor $k$, all data collapse onto the relation
\begin{equation}
    \mathrm{Ca}_c=k\phi^{-1}, \label{eq:Cac}
\end{equation}
including those obtained at different bulk viscosities.

\begin{figure}[htbp]
\begin{center}
\includegraphics[width=8.5cm]{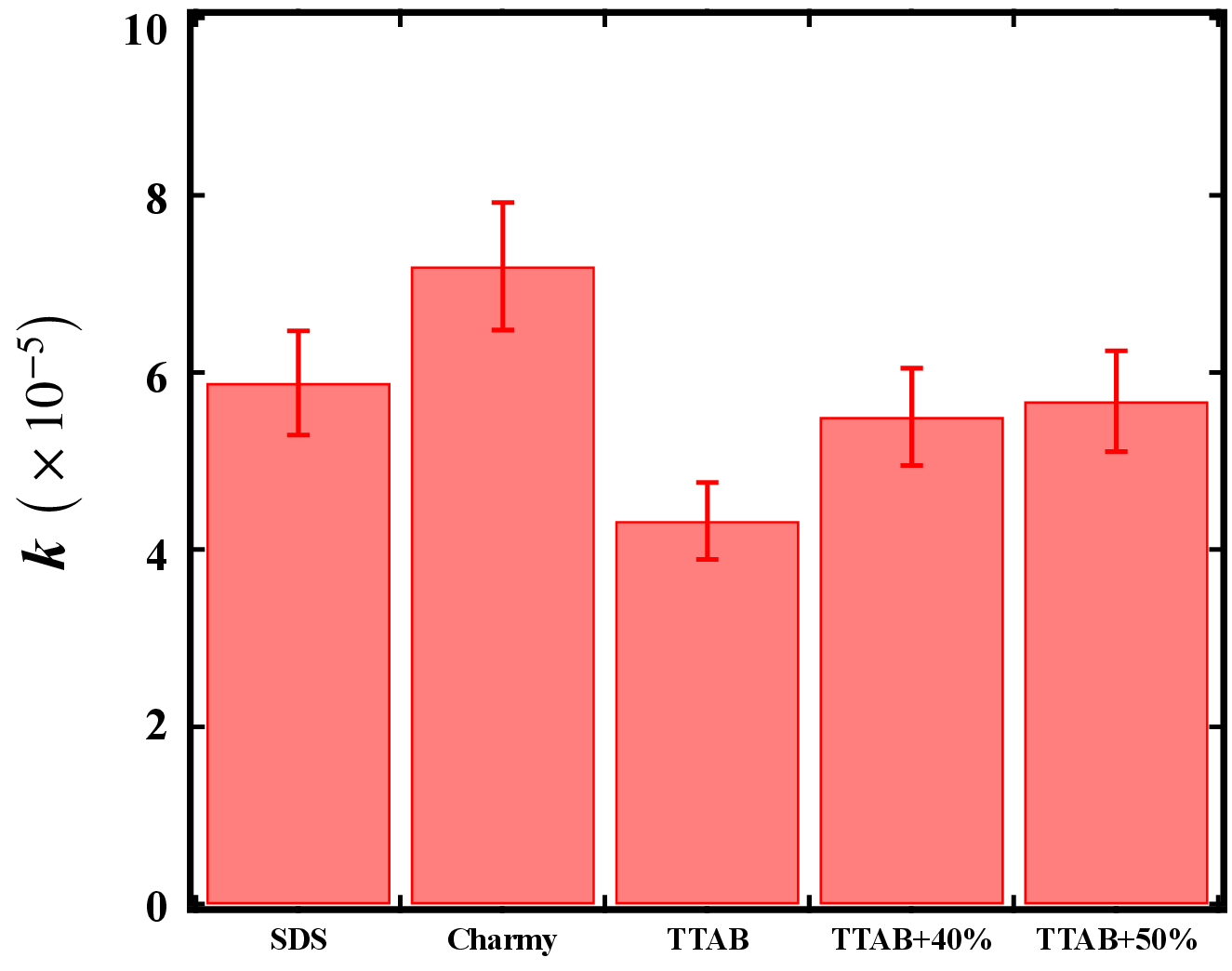}
\caption{
Prefactor $k$ obtained from $\mathrm{Ca}_c = k\phi^{-1}$ for each solution. 
The values of $k$ remain of the same order for all surfactant systems, while a systematic variation is observed with glycerol concentration in the TTAB solutions.
Error bars represent the uncertainty obtained from the fitting.}
\label{k}
\end{center}
\end{figure}

The values of the prefactor $k$ are summarized in Fig.~\ref{k}. They range from $4.6\times10^{-5}$ to $7.0\times10^{-5}$ and are therefore of the same order of magnitude for all solutions. Nevertheless, $k$ varies with glycerol concentration even within the TTAB system. The possible origin of this variation is discussed in Sec.~\ref{re5}.

\subsection{Model} \label{re5}
We next consider the physical mechanism that determines the transition from the slip regime to the scraping regime. During scraping, the relative
motion between the bubble structure and the liquid in the Plateau border network generates a viscous stress. The characteristic velocity gradient is
estimated as $U/r$, where $U$ is the scraping velocity and $r$ is the characteristic radius of the Plateau border. The corresponding viscous stress
is therefore
\begin{equation}
    \sigma_{\mathrm{vis}}
    \sim
    \eta\frac{U}{r},
\end{equation}
where $\eta$ is the liquid viscosity.

We assume that this stress acts over a cross-sectional area of order $r^2$.
The force generated through the Plateau-border network is then transmitted
over a bubble-scale distance of order $a$, where $a$ is the characteristic
bubble radius. The viscous work available for inducing a neighboring
rearrangement is consequently estimated as
\begin{equation}
    W_{\mathrm{vis}}
    \sim
    C_W
    \left(
        \eta\frac{U}{r}
    \right)
    r^2a
    =
    C_W\eta Ura.
    \label{eq:viscous_work}
\end{equation}
Here, $C_W$ is a dimensionless coefficient describing the efficiency with which viscous stresses in the Plateau border network are transmitted to the bubble structure. 
Its value is expected to depend on the mobility of the interface characterized by the Boussinesq number $\mathrm{Bo} = \eta_s/\eta r$~\cite{cantat2013}. 
A large $\mathrm{Bo}$ corresponds to a relatively immobile interface, whereas a smaller $\mathrm{Bo}$ corresponds to a more mobile, or softer, interface.

At the transition, the viscous work is assumed to become comparable to the effective energetic cost required for the propagation of a T1 event,
\begin{equation}
    W_{\mathrm{vis}}
    \sim
    E_{\mathrm{T1}}^{\mathrm{eff}}.
    \label{eq:energy_balance}
\end{equation}
The superscript ``eff'' emphasizes that this quantity is not the equilibrium energy barrier of an isolated T1 event. 
It represents the energetic cost for one rearrangement to induce a subsequent rearrangement under the externally driven and spatially asymmetric conditions produced by scraping.

Using the definition $Ca = \eta U/\gamma$, where $\gamma$ is the surface tension, Eqs.~(\ref{eq:viscous_work}) and (\ref{eq:energy_balance}) give, apart from a dimensionless prefactor,
\begin{equation}
    E_{\mathrm{T1}}^{\mathrm{eff}}
    \sim
    \gamma \mathrm{Ca_c} ra,
    \label{eq:ET1_general}
\end{equation}
where $\mathrm{Ca}_c$ is the critical capillary number. For a dry foam, the characteristic Plateau-border radius scales as
\begin{equation}
    r\sim a\phi^{1/2},
    \label{eq:r_phi}
\end{equation}
where $\phi$ is the liquid fraction. Combining Eq.~(\ref{eq:r_phi}) with the experimentally observed relation
$\mathrm{Ca}_c\sim\phi^{-1}$ yields
\begin{equation}
    E_{\mathrm{T1}}^{\mathrm{eff}}
    \sim\gamma a^2\phi^{-1/2}.
\end{equation}
yields
\begin{equation}
    E_{\mathrm{T1}}^{\mathrm{eff}}
    \sim
    \gamma a^2\phi^{-1/2}.
    \label{eq:ET1_phi}
\end{equation}
Thus, the experimental scaling directly determines the liquid-fraction dependence of the effective energetic cost for T1 propagation. 
In particular, the energy required to propagate rearrangement activity increases as the foam becomes drier.

Using $r\sim a\phi^{1/2}$, Eq.~(\ref{eq:ET1_phi}) can equivalently be written as
\begin{equation}
    E_{\mathrm{T1}}^{\mathrm{eff}}
    \sim
    \frac{\gamma}{r}a^3.
    \label{eq:ET1_PV_scaling}
\end{equation}
Because $\gamma/r$ represents the characteristic Laplace pressure, this scaling has the form of pressure--volume work. We therefore write
\begin{equation}
    E_{\mathrm{T1}}^{\mathrm{eff}}
    \sim
    P_{\mathrm{L}}\Delta V_{\mathrm{eff}},
    \qquad
    P_{\mathrm{L}}\sim\frac{\gamma}{r},
    \qquad
    \Delta V_{\mathrm{eff}}=C_Va^3,
    \label{eq:ET1_PV}
\end{equation}
where $C_V$ is a dimensionless measure of the effective volumetric deformation associated with T1 propagation, and the experimental prefactor $k$ gives $k = C_V/C_W$. 
Thus, the relation 
\begin{eqnarray}
E_{\mathrm{T1}}^{\mathrm{eff}} \propto \frac{\gamma}{r}a^3,
\end{eqnarray}
suggests that a small local compression of the bubbles is a governing parameter for a T1 event under scraping.  

The prefactor $k$ remains of the same order for TTAB, SDS, and Charmy. This is consistent with the similar Boussinesq numbers reported for these surfactant systems, suggesting that their efficiencies of viscous-stress transmission, represented by $C_W$, are also comparable. 
A systematic change in $k$ is nevertheless observed when glycerol is added to the TTAB solution. The bulk viscosity $\eta$ then increases substantially,
whereas the interfacial viscosity $\eta_s$ is not expected to increase by the same factor. The resulting decrease in $\mathrm{Bo}$ makes the interface more mobile and is expected to reduce $C_W$. This interpretation is consistent with the observed increase in $k$ with glycerol concentration.

\subsection{Local compression as a trigger for T1 propagation} 

The experimentally observed scaling suggests that the propagation of a T1 event involves a small compressive component. If $C_W$ is of order unity, the estimated volume change is 
\begin{equation}
    \frac{\Delta V_{\mathrm{eff}}}{a^3}
    \sim 6.0 \times10^{-5}.
\end{equation}
This corresponds to a characteristic linear scale
\begin{equation}
    \frac{\delta l}{a}
    \sim
    \left(
    \frac{\Delta V_{\mathrm{eff}}}{a^3}
    \right)^{1/3}
    \sim 3.9 \times10^{-2},
\end{equation}
so that the inferred local deformation is of the order of a few percent of the bubble size.

To obtain an order-of-magnitude comparison, we estimate the transient increase in surface-energy cost as
\begin{equation}
    E_A \sim \gamma\Delta A \sim \gamma \delta l^2
    \sim 1.5 \times10^{-3}\gamma a^2.
\end{equation}
By contrast, the compressive work is
\begin{equation}
    E_V
    \sim
    \Delta P_L\Delta V_{\mathrm{eff}}
    \sim
    \frac{\gamma}{r}\Delta V_{\mathrm{eff}}
    \sim
    6.0 \times10^{-5}
    \frac{a}{r}
    \gamma a^2.
\end{equation}
Using $r/a \sim 1.7 \phi^{1/2}$~\cite{Weaire2001, cantat2013} and $\phi \simeq 0.1$, we obtain
\begin{equation}
    E_V \sim 1.1 \times10^{-4}\gamma a^2,
\end{equation}
which is approximately one order of magnitude smaller than $E_A$. 
This order-of-magnitude estimate indicates that the compressive pathway can be energetically more favorable than a volume-preserving deformation accompanied by a comparable transient increase in interfacial area.

This interpretation is also plausible for a jammed foam. Even a small local volume change can generate an elastic heterogeneity in the otherwise rigid bubble network and thereby facilitate localized bubble motion. 
The local compression may therefore act as a trigger for the propagation of the next T1 event.

\section{Conclusion}

Foam scraping exhibits a dynamic transition between a slip state, in which the foam passes beneath the plate, and a scraping state, in which it is expelled from the confined region~\cite{Endo2023}. 
Previous studies have shown that this transition cannot be explained solely by the mean shear rate or the macroscopic yield stress and have suggested that it is controlled by the propagation of local T1 rearrangements~\cite{Endo2025}. 
However, the physical parameter governing the propagation of these rearrangements has remained unclear.
To address this problem, we examined the effects of liquid fraction, bulk viscosity, and surfactant system on the critical scraping velocity. 
For all conditions examined, the critical capillary number followed $\mathrm{Ca}_c = k\phi^{-1}$, where $k$ is a solution-dependent prefactor.
This scaling demonstrates that the interstitial liquid plays an active role in determining whether a local rearrangement propagates through the foam.

We proposed a local fluid--structure-coupling model in which viscous work generated in the Plateau-border network is transmitted to the bubble structure and competes with the effective energetic cost required for one T1 event to trigger the next. 
The experimentally observed scaling implies that this cost is governed by the Laplace pressure and is consistent with a small local compressive deformation near the rearranging region. 
Such local compression may generate an elastic heterogeneity in the jammed bubble network and facilitate the propagation of subsequent T1 events.

These results identify local coupling between interstitial flow and bubble deformation as a mechanism controlling the macroscopic slip--scraping
transition. Because deformation in emulsions, biological tissues, and other soft jammed materials is also mediated by localized rearrangements, the present framework may provide a broader basis for understanding how fluid--structure coupling selects macroscopic deformation modes in multicomponent soft matter systems.

\section*{Acknowledgments} 
R. K. was supported by a Grant-in-Aid for Scientific Research (B) (Grant No. 26K00676) from the JSPS.

\section*{Author contributions}
RK conceived the project, and ME performed the experiments.
RK developed the model for the transition mechanism and wrote the manuscript.

\section*{Data availability}
All data supporting the findings of this study are included within the article.

\section*{Competing interests} 
The authors declare that they have no competing financial interests. 

\section*{Correspondence} 
Correspondence and requests for materials should be addressed to R.K. 
(kurita@tmu.ac.jp).

\bibliography{Foam}

@article{Yanagishima2021,
	author = {Yanagishima, Taiki and Liu, Yanyan and Tanaka, Hajime and Dullens, Roel P. A.},
	issue = {2},
	journal = {Phys. Rev. X},
	month = {Jun},
	numpages = {17},
	pages = {021056},
	title = {Particle-Level Visualization of Hydrodynamic and Frictional Couplings in Dense Suspensions of Spherical Colloids},
	volume = {11},
	year = {2021}}

@article{Furukawa2006,
	author = {Furukawa, Akira and Tanaka, Hajime},
	date = {2006/09/01},
	id = {Furukawa2006},
	journal = {Nature},
	number = {7110},
	pages = {434--438},
	title = {Violation of the incompressibility of liquid by simple shear flow},
	volume = {443},
	year = {2006}}

@article{Araki2008,
	author = {Araki, Takeaki and Tanaka, Hajime},
	journal = {Progress of Theoretical Physics Supplement},
	month = {05},
	pages = {37-46},
	title = {Dynamics of Colloidal Particles in Soft Matters},
	volume = {175},
	year = {2008}}

@article{Tateno2025,
	author = {Michio Tateno and Jiaxing Yuan and Hajime Tanaka},
	journal = {Journal of Colloid and Interface Science},
	pages = {21-28},
	title = {The impact of colloid-solvent dynamic coupling on the coarsening rate of colloidal phase separation},
	volume = {684},
	year = {2025}}

@article{Kaneda2025,
	author = {Aoi Kaneda and Rei Kurita},
	journal = {J. Colloid and Interface Sci.},
	pages = {137746},
	title = {Absorptive limits of foams governed by kinematic coupling between solution and bubbles},
	volume = {695},
	year = {2025}}

@article{Endo2025,
	author = {Masaya Endo and Rei Kurita},
	journal = {Phys. Rev. Res.},
	pages = {023013},
	title = {Critical-like behavior in foam dynamics: Transition from slip to scraping},
	volume = {7},
	year = {2025}}

@article{Endo2023,
	author = {Masaya Endo and Marie Tani and Rei Kurita},
	journal = {J. Colloid and Interface Sci.},
	pages = {1612-1618},
	title = {Scraping of foam on a substrate},
	volume = {650},
	year = {2023}}

@article{Deblais2015,
	author = {A. Deblais and R. Harich and D. Bonn and A. Colin and H. Kellay},
	journal = {Langmuir},
	pages = {5971--5981},
	title = {Spreading of an Oil-in-Water Emulsion on a Glass Plate: Phase Inversion and Pattern Formation},
	volume = {31},
	year = {2015}}

@article{Katgert2013,
	author = {Gijs Katgert and Brian P. Tighe and Martin van Hecke},
	journal = {Soft Matter},
	pages = {9739-9746},
	title = {The jamming perspective on wet foams},
	volume = {9},
	year = {2013}}

@article{Merrer2012,
	author = {M. L. Merrer and S. Cohen-Addad and R. H{\"o}hler},
	journal = {Phys. Rev. Lett.},
	pages = {188301},
	title = {Bubble Rearrangement Duration in Foams near the Jamming Point},
	volume = {108},
	year = {2012}}

@article{Danov2014,
	author = {K. D. Danov and P. A. Kralchevsky and K. P. Ananthapadmanabhan},
	journal = {Adv. Colloid Interface Sci.},
	pages = {17--45},
	title = {Micelle--monomer equilibria in solutions of ionic surfactants and in ionic--nonionic mixtures: A generalized phase separation model},
	volume = {206},
	year = {2014}}

@book{Weaire2001,
	author = {Weaire, Denis L and Hutzler, Stefan},
	publisher = {Oxford University Press},
	title = {The physics of foams},
	year = {2001}}

@book{cantat2013,
	author = {Cantat, Isabelle and Cohen-Addad, Sylvie and Elias, Florence and Graner, Fran{\c{c}}ois and H\"{o}hler, Reinhard and Pitois, Olivier and Rouyer, Florence and Saint-Jalmes, Arnaud},
	publisher = {OUP Oxford},
	title = {Foams: structure and dynamics},
	year = {2013}}

@article{cohen2013,
	author = {Cohen-Addad, Sylvie and H\"{o}hler, Reinhard and Pitois, Olivier},
	journal = {Annual Review of Fluid Mechanics},
	pages = {241--267},
	publisher = {Annual Reviews},
	title = {Flow in foams and flowing foams},
	volume = {45},
	year = {2013}}

@article{Furuta2016,
	author = {Furuta, Yujiro and Oikawa, Noriko and Kurita, Rei},
	journal = {Sci. Rep.},
	number = {1},
	pages = {37506},
	publisher = {Nature Publishing Group},
	title = {Close relationship between a dry-wet transition and a bubble rearrangement in two-dimensional foam},
	volume = {6},
	year = {2016}}

@article{Kurita2017,
	author = {Kurita, Rei and Furuta, Yujiro and Yanagisawa, Naoya and Oikawa, Noriko},
	journal = {Physical Review E},
	number = {6},
	pages = {062613},
	publisher = {APS},
	title = {Dynamical transition in a jammed state of a quasi-two-dimensional foam},
	volume = {95},
	year = {2017}}

@article{marchand2020,
	author = {Marchand, Manon and Restagno, Fr{\'e}d{\'e}ric and Rio, Emmanuelle and Boulogne, Fran{\c{c}}ois},
	journal = {Phys. Rev. Lett.},
	number = {11},
	pages = {118003},
	publisher = {APS},
	title = {Roughness-Induced Friction on Liquid Foams},
	volume = {124},
	year = {2020}}

@article{siemens2010,
	author = {Siemens, Alexander ON and Van Hecke, Martin},
	journal = {Physica A: Statistical Mechanics and its Applications},
	number = {20},
	pages = {4255--4264},
	publisher = {Elsevier},
	title = {Jamming: A simple introduction},
	volume = {389},
	year = {2010}}

@article{Sano2016,
	author = {Sano, Masaki and Tamai, Keiichi},
	journal = {Nat. Phys.},
	number = {3},
	pages = {249--253},
	publisher = {Nature Publishing Group UK London},
	title = {A universal transition to turbulence in channel flow},
	volume = {12},
	year = {2016}}

@article{Haye2000,
	author = {Hinrichsen, Haye},
	journal = {Adv. in Phys.},
	number = {7},
	pages = {815--958},
	publisher = {Taylor \& Francis},
	title = {Non-equilibrium critical phenomena and phase transitions into absorbing states},
	volume = {49},
	year = {2000}}

@article{Takeuchi2009,
	author = {Takeuchi, Kazumasa A and Kuroda, Masafumi and Chat{\'e}, Hugues and Sano, Masaki},
	journal = {Phys. Rev. E},
	number = {5},
	pages = {051116},
	publisher = {APS},
	title = {Experimental realization of directed percolation criticality in turbulent liquid crystals},
	volume = {80},
	year = {2009}}

\clearpage

\end{document}